\documentclass[pra,aps,twocolumn,showpacs,tightenlines]{revtex4}
\usepackage{graphics,bm}
\usepackage{graphicx}
\newcommand{\beq}{\begin{equation}}
\newcommand{\eeq}{\end{equation}}
\newcommand{\bqa}{\begin{eqnarray}}
\newcommand{\eqa}{\end{eqnarray}}

\def\square{\vcenter{\vbox{\hrule height.4pt
          \hbox{\vrule width.4pt height8pt
          \kern8pt\vrule width.4pt}\hrule height.4pt}}}

\begin{document}

\title{Effective Field Theory for Goldstone Bosons in Nonrelativistic 
Superfluids}

\author{Jens O. Andersen}
\email{andersen@phys.uu.nl}
\affiliation{Institute for Theoretical Physics,
         University of Utrecht, Leuvenlaan 4,
         3584 CE Utrecht, The Netherlands}
\date{\today}

\begin{abstract}
We consider nonrelativistic superfluids where the global $U(1)$-symmetry is
spontaneously broken. At sufficiently long wavelengths, the relevant 
degree of freedom is the massless Goldstone mode and
we construct an effective low energy theory for the
Goldstone boson. The damping rate of collective excitations
at low energy is calculated.
In the case of a weakly interacting Bose gas, we recover 
the results by Beliaev, and by Hohenberg and Martin.
\end{abstract}

\preprint{ITF-UU-02/xx}
\maketitle

\section{Introduction}

Spontaneous symmetry breaking is central in our description of many phenomena
in condensed matter and high-energy physics (see e.g. Ref.~\cite{ssb}
and references therein). 
An example of spontaneous symmetry breaking
is the breaking of the global $U(1)$-symmetry in $^4$He and the appearance of
superfluidity below a certain critical temperature.
Associated with the breaking of the symmetry, there is a nonzero value of
an order parameter and a condensate of particles in the zero-momentum
state.
When a global continuous symmetry is broken, the Goldstone 
theorem~\cite{gold} states
that there is a gapless excitation for each generator that does not leave the
ground state invariant. In the case of nonrelativistic Bose gases, one
identifies the Goldstone mode with the phonons.

At long wavelengths where the relevant degrees of freedom are the massless
Goldstone bosons, one would naturally like to have an effective low energy 
theory that contains only these degrees of freedom.
Such an effective field theory may simplify calculations of the long-wavelength
properties of the system and one can obtain 
model independent predictions with a minimum of assumptions.

An example of a low energy theory for massless Goldstone modes
is chiral perturbation theory in QCD~\cite{chiral}. 
In massless QCD, chiral symmetry is broken and one identifies
the Goldstone bosons with the pions. However, QCD is a confining and
strongly interacting theory at low energies. 
Thus 
the coefficients of the chiral Lagrangian cannot be calculated
from QCD. Instead, the coefficients of the effective Lagrangian are fixed
by experiments. In other cases, one can determine the coefficients
of the low-energy theory as functions of the coupling constants in the
underlying theory e.g.
by a matching procedure. One calculates physical
quantities at low energies perturbatively
and demand they be the same in the 
full and in the effective theory. The coefficients of the effective
theory then encode the physics at short distances. Nonrelativistic 
QED~\cite{nrqed} is an example of such an effective field theory that is
taylored to perform low-energy (bound state) calculations, where the 
coefficients encode the effects of relativistic momenta.

In this paper, we construct an effective low energy theory for the Goldstone
boson in nonrelativistic superfluids with a broken global
$U(1)$-symmetry. The
effective Lagrangian allows one to calculate physical quantities in the
long-wavelength limit with ease without assuming weak coupling.

\section{Effective Lagrangian}
The effective theory for the Goldstone field $\phi$ can be constructed
using the methods of effective field theory~\cite{eft}. 
Once the symmetries of the theory have been identified, one writes down the 
most general Lagrangian ${\cal L}_{\rm eff}$
that is consistent with these symmetries. Examples of symmetries are
time-reversal invariance and Galilean invariance, and these symmetries
severely restrict the possible terms in ${\cal L}_{\rm eff}$.
For instance, Galilean invariance implies that the terms in the effective
theory are powers of the combination 
$i\partial_{\tau}\phi-{1\over2}\left(\nabla\phi\right)^2$ 
($\hbar=k_B=m=1$ henceforth).
The Euclidean Lagrangian can then be written as
\bqa\nonumber
{\cal L}_{\rm eff}&=&
-c_1\!\left[
i\partial_{\tau}\phi-{1\over2}\left(\nabla\phi\right)^2
\right]
-{1\over2}c_2\!\left[i\partial_{\tau}\phi-{1\over2}
\left(\nabla\phi\right)^2\right]^2
\\&&
+d_1\!\left[\partial_{\tau}\phi-{1\over2}
\left(\nabla\phi\right)^2\right]^3
+\delta{\cal L}_{\rm eff}\;,
\label{efflag}
\eqa
where $c_1, c_2,...$ are parameters that must be determined either by 
experiment or by matching.
The first term is a total derivative and can be omitted if one is not
interested in topologically nontrivial field configuration such as vortices.
The term $\delta{\cal L}_{\rm eff}$ contains all terms that are higher order
in the field $\phi$ and derivatives thereof. 
We have also omitted the unit operator $f$~\cite{eric}, 
which is necessary to include if one is interested in the equation of state. 
The unit operator $f$ can be interpreted as the contribution to the
free energy from large momenta.  
We will briefly discuss this issue below.
The terms in Eq.~(\ref{efflag}) that we have shown explicitly, gives rise to 
a linear dispersion relation
for the Goldstone bosons, and the effective Lagrangian is therefore
valid only for momenta where this is a good approximation.
At larger momenta, the dispersion relation is no longer linear and 
terms of the form $(\nabla^2\phi)^2$ must be included.
The scale at which the effective Lagrangian no longer can be applied is
given by the scale where other degrees 
of freedom than the Goldstone boson become important.
In the case of a weakly interacting Bose gas, this scale is given by the
coherence length $\xi$.

The free propagator that corresponds to the Lagrangian~(\ref{efflag}) is
\bqa
\Delta(p_0,{\bf p})&=&{1\over c_1p_0^2+c_2p^2}\;,
\eqa
where $p_0=2\pi n T$ are the Matsubara frequencies and $p=|{\bf p}|$.
The dispersion relation is given by the pole of the 
Minkowski space propagator:
\bqa
\epsilon({\bf p})&=&\sqrt{{c_2\over c_1}}p\;.
\eqa
The nonrelativistic hydrodynamic
speed of sound $c$ is then given by $\sqrt{c_2/c_1}$. In some cases,
one does not know the underlying microscopic theory, and so one cannot
determine the couplings $c_1, c_2,...$ by matching. Instead, one performs
experiments at low energies. For instance, the ratio $c_2/c_1$ is fixed
by measuring the speed of sound. 

In the case of a weakly interacting Bose
gas, we know the underlying theory and we can calculate the coefficients
$c_1, c_2,...$ in terms of the parameters of the full theory.
The Euclidean action for nonrelativistic bosons at low energy is
\bqa\nonumber
S[\psi^{\dagger},\psi]&=&
\int_0^{\beta}\!d\tau\!
\int\!d{\bf x}\;
\psi^{\dagger}
\Bigg[
{\partial\over\partial\tau}
-{1\over2}\nabla^2-\mu\Bigg]\psi\\
&-&{1\over2}\int_0^{\beta}d\tau \int d{\bf x}\;
g(\psi^{\dagger}\psi)^2
\label{ac}\;,
\eqa
where $\psi({\bf x})$ annihilates a boson at position ${\bf x}$, $\mu$
is the chemical potential, and $g=4\pi a$, where $a$ is the $s$-wave
scattering length. If we substitute 
$\psi({\bf x},\tau)=\sqrt{n_0+\sigma({\bf x},\tau)}e^{i\phi({\bf x},\tau)}$ 
into Eq.~(\ref{ac}), we obtain the action
\bqa\nonumber
S[\sigma,\phi]
&=&\int_0^{\beta}\!d\tau\!\int d{\bf x}\!
\left\{[n_0g-\mu]n_0\sigma+
{1\over2}i\left[\sigma{\partial\phi\over\partial\tau}
-\phi{\partial\sigma\over\partial\tau}
\right]\right.
\\&&\nonumber
\left.
+{1\over2}n_0\left(\nabla\phi\right)^2
+{1\over2}\sigma\left[{1\over4n_0}\nabla^2+V_0\right]
\sigma
\right.
\\
&&
\left.+{1\over2}\sigma\left(\nabla\phi\right)^2+...\right\}\;,
\label{x100}
\eqa
where we have dropped total derivatives as well as an infinite
series of higher order momentum-dependent interactions.
By using the classical equation of motion for $\sigma({\bf x},\tau)$,
we obtain from Eq.~(\ref{x100}) the following action for the phase $\phi$:
\bqa\hspace{-.1cm}
S[\phi]&=&\int_0^{\beta}\!d\tau\!\int d{\bf x}\!
\left[{1\over2g}
(\partial_{\tau}\phi)^2
+{1\over 2}n_0\left(\nabla\phi\right)^2
+...\right] .
\label{pact}
\eqa
Using Eq.~(\ref{efflag}) and
$S=\int_0^{\beta}d\tau\int d {\bf x}\;{\cal L}_{\rm eff}$, and comparing with
Eq.~(\ref{pact}) 
we find the coefficients $c_1=n_0$ and $c_2=1/g$. 

As an application of the effective low-energy Lagrangian~(\ref{efflag})
for the Goldstone bosons,
we next calculate the damping rate
of a long-wavelength excitation in the superfluid. 
The propagator has poles in the complex energy plane, and the real part gives 
the energy $\epsilon(p)$ of the excitation, 
while the imaginary part gives the damping rate.

\begin{figure}
\includegraphics[width=8cm]{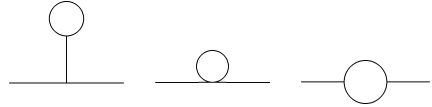}
\vspace{-9cm}
\caption{\label{dia}
One-loop 
Feynman diagrams for the self-energy $\Pi(p_0,{\bf p})$.}
\end{figure}

The effective Lagrangian~(\ref{efflag}) gives rise to three-point and 
four-point interactions and the one-loop
Feynman diagrams contributing to the self-energy $\Pi(p_0,{\bf p})$
are shown in Fig.~\ref{dia}. 
The first diagram vanishes identically
since the vertices are momentum dependent. The second and third diagrams are 
ultraviolet divergent and must be regularized.
In this paper, we use dimensional regularization 
to regulate both infrared and ultraviolet divergences.
In dimensional regularization, one calculates the loop
integrals in $d=3-2\epsilon$ dimensions for values of $\epsilon$ where
the integrals converge. One then analytically continues back to $d=3$
dimensions.
With dimensional regularization, an arbitrary renormalization scale $M$
is introduced. An advantage of dimensional regularization is that it 
automatically sets power divergences to zero, while logarithmic divergences
show up as poles in $\epsilon$. The second diagram is proportional to
$p^2$ and has pentic, cubic, and linear divergences 
in the ultraviolet. It is therefore set to
zero with dimensional regularization. The third graph has a logarithmic
ultraviolet divergence. The one-loop self-energy is
\bqa\nonumber
\Pi(p_0,{\bf p})&=&{1\over4}c_2^2
\left({e^{\gamma}M^2\over4\pi}\right)^{\epsilon}
\sum_{k_0}\!\int{d^dk\over(2\pi)^d}\Bigg\{
\\ \nonumber
&&
{1\over[c_1k_0^2+c_2k^2][c_1(p_0+k_0)^2+c_2({\bf p}+{\bf k})^2]}
\\ \nonumber
&&\times\Big[
4\left({\bf p}\!\cdot\!{\bf k}\right)^2(p_0+k_0)^2
+(p_0k^2+k_0p^2)^2
\\
&&+
4({\bf p}\cdot{\bf k})(p_0+k_0)(p_0k^2+k_0p^2)
\Big]\Bigg\}\;,
\eqa
where $\gamma\approx0.5772$ is the Euler-Mascharoni constant.
In the zero-temperature limit, the sum over 
Matsubara frequencies becomes
an integral over the Euclidean energy. The next step in 
evaluating the self-energy is to introduce a Feynman parameter $y$.
After integrating 
over energy, momentum, and finally over $y$,
the self-energy can be expanded in powers of $\epsilon$. The result is
\bqa\nonumber
\Pi(p_0,{\bf p})&=&-{p_0^2\over320\pi^2\sqrt{c_1c_2}}
\left[49p^4+30{c_1\over c_2}p_0^2p^2+5{c_1^2\over c_2^2}p_0^4\right]
\\&&
\times\left[
{1\over\epsilon}+\log\left({M^2\over p_0^2+p^2}\right)+g(p_0,p)\right]\;,
\eqa
where $g(p_0,p)$ is an analytic function that is not important to us.
After analytic continuation to real frequencies, 
$p_0\rightarrow i\omega+\eta$, we find the
the damping rate 
$\gamma={{\rm Im}\Pi(-i\epsilon({\bf p})+\eta,{\bf p})}/2\epsilon({\bf p})$
\bqa
\gamma&=&{3p^5\over640\pi c_1}\;.
\label{gammaz}
\eqa
Note that the result is independent of the coupling $c_2$.
By measuring the damping of long-wavelength excitation one can determine
$c_1$, while measuring the speed of sound $c$ will determine 
ratio $c_1/c_2$. Using the value $c_1=n_0$ for a weakly interacting Bose
gas, we recover the result by Beliaev~\cite{beli}
(See also Refs.~\cite{w1,w2}).
 
The quadractic part of action~(\ref{pact}) has 
recently been used a starting point for calculating
 phase fluctuations in trapped Bose gases
at low temperature~\cite{1d,2d}. 
The effective field theory approach clearly shows
why such an approach is incorrect. The phase-alone action is only valid
for momenta less than the inverse
coherence length, while in Refs.~\cite{2d,1d},
it was used for all momenta.
The correct treatment of the phase fluctuations
was given in Refs.~\cite{jh,paper}.

Consider next the pressure for a dilute
Bose gas at zero temperature. If one calculates 
${\cal P}$ using the effective Lagrangian~(\ref{efflag}), one
gets zero. This disagrees with the one-loop result
of Lee and Yang~\cite{leeyang}. The reason is simply that one must include
the unit operator $f$ in the effective Lagrangian~(\ref{efflag}) 
and that contribution to the pressure exactly equals the standard result.

\section{Finite temperature effects}
We next consider finite-temperature effects using the effective low-energy
Lagrangian~(\ref{efflag})

The pressure at one-loop order is given by
\bqa\nonumber
{\cal P}&=&-{1\over2}\left({e^{\gamma}M^2\over4\pi}\right)^{\epsilon}
\sum_{k_0}\!\int{d^dk\over(2\pi)^d}
\log\left[c_1k_0^2+c_2k^2\right]\\
&=&{\pi^2T^4\over90}\left({c_1\over c_2}\right)^{3/2}\;.
\eqa
For a weakly interacting Bose gas, 
this result is in agreement with the leading  temperature-dependent term
in a low-temperature expansion carried out by Lee and Yang~\cite{leeyang2}
(See also Ref.~\cite{finn2}).
This simply reflects the fact that the massless Goldstone
modes are the important thermal
excitations at low temperature.

Finally we consider damping at finite $T$. The imaginary part of the
self-energy can be calculated in a similar manner as before, except that
the integral over Euclidean energy is replaced by a summation over Matsubara
frequencies. The result is
\bqa
\gamma&=&{3\pi^3pT^4\over40c_1}\;.
\label{ftgamma}
\eqa
In the case of a weakly interacting Bose gas, this results was first 
obtained by Hohenberg and Martin~\cite{maho}. 
Using the Lagrangian~(\ref{x100}), Liu~\cite{w1} obtained a more general
result to lowest order in $p$.
This result is valid for $p\ll T$ and $p\ll n_0g$ 
and reduces to $3\pi^2pT^4/(40n_0)$ 
in the limit $T/n_0g\rightarrow 0$.

\section{Discussion}
In this paper, we have constructed an effective field theory for Goldstone
modes at low energies for a nonrelativistic superfluid with a broken 
$U(1)$-symmetry. The results for the damping of collective excitations 
is in agreement with those of Beliaev, and by Hohenberg and Martin
for the weakly interacting theory. However, the results go beyond
because nowhere have we assumed weak coupling. The only assumptions
are Galilean invariance and low energy. Hence, 
the results~(\ref{gammaz}) and~(\ref{ftgamma}) are valid for {\it any}
nonrelativistic superfluid with a broken $U(1)$-symmetry in the long-wavelength
limit.

There is another way of arrriving at a low-energy theory for the Goldstone
modes,
namely by calculating the quantum effective action~\cite{ssb}.
In the effective action approach,
integrating out ``short-distance physics'' amounts to minimizing
with respect to the amplitude of the field $\psi({\bf x},\tau)$.
Such an approach was used in Ref.~\cite{damn} to calculate a low-energy
effective Lagrangian for relativistic superfluids where the $U(1)$
baryon symmetry is spontaneously broken. 
In Ref.~\cite{damn}, it was shown that 
the scattering amplitudes among the Goldstone bosons can be found
once the equation of state is known.
Applying the formalism to the dilute Bose 
calculating the quantum effective action and using ${\cal P}(\mu)=\mu^2/2g$,
one would immediately obtain Eq.~(\ref{efflag}) with the coefficients
$c_1=\mu/g=n_0$ and $c_2=1/g$.

\section*{Acknowledgments}
This work was supported by the Stichting voor
Fundamenteel Onderzoek der Materie
(FOM), which is supported by the Nederlandse Organisatie voor Wetenschappelijk
Onderzoek (NWO).


\begin{thebibliography}{99}
\bibitem{ssb} S. Weinberg, {\it The Quantum Theory of Fields II, 
Modern Applications} (Cambridge University Press, Cambridge England 1996).
\bibitem{gold} J. Goldstone, Nuovo Cim. {\bf 19}, 154 (1961).
\bibitem{chiral}J. Gasser and H. Leutwyler, Annals Phys. {\bf 158}, 142 
(1984); Nucl. Phys. {\bf B 250}, 465 (1985).
\bibitem{nrqed} W.E. Caswell and G.P. Lepage, Phys. Lett. {\bf B167}, 
437 (1986).

\bibitem{eft}H. Georgi, Ann. Rev. Nucl. Part. Sci. {\bf 43}, 209 (1993). 

\bibitem{eric}E. Braaten, A. Nieto, Phys. Rev. D {\bf 51}, 6990 (1995).

\bibitem{beli}S.T. Beliaev, Sov. J. Phys. {\bf 7}, 289 (1958);
{\bf 34}, 299 (1958).

\bibitem{w1}W.V. Liu, Phys. Rev. Lett. {\bf 79}, 4056 (1997).
\bibitem{w2}W.V. Liu, Int. J. Mod. Phys. B {\bf 12}, 2103 (1998).


\bibitem{2d} D. S. Petrov, M. Holzmann, and G.V. Shlyapnikov,
Phys. Rev. Lett. {\bf 84}, 2551 (2000).

\bibitem{1d} D. S. Petrov, G.V. Shlyapnikov, and J.T.M. Walraven,
Phys. Rev. Lett. {\bf 85}, 3745 (2000).

\bibitem{jh}J. O. Andersen, U. Al Khawaja, 
and H. T. C. Stoof, Phys. Rev. Lett {\bf 88}, 070407 (2002). 
\bibitem{paper}
U. Al Khawaja, J.O. Andersen, N.P. Proukakis, and H.T.C. Stoof,
Phys. Rev. A {\bf 66}, 013615 (2002).


\bibitem{leeyang}T.D. Lee and C.N. Yang, Phys. Rev. {\bf 105}, 1119 (1957).
\bibitem{leeyang2}
T.D. Lee and C.N. Yang, Phys. Rev. {\bf 112}, 1419 (1958).
\bibitem{finn2}T. Haugset, H. Haugerud, and F. Ravndal, 
Ann. Phys. (N.Y) {\bf 226}, 27 (1998).

\bibitem{maho}P.C. Hohenberg and P.C. Martin, Ann Phys. (N.Y.) {\bf 34},
291 (1965).
\bibitem{damn} D.T. Son, hep-ph/0204199.





\end{thebibliography}
\end{document}